\documentclass[12pt]{article}  % commençé le 20 nov. 06
\def\rv{{\bf r}}
\def\vv{{\bf v}}

\def\bv{{\bf b}}

\def\kv{{\bf k}}
\def\jv{{\bf j}}
\def\sv{{\bf s}}
\def\Sv{{\bf S}}
\def\Ev{{\bf E}}

\def\Bv{{\bf B}}
\def\Av{{\bf A}}
\def\Jv{{\bf J}}

\def\Pv{{\bf P}}
\def\Qv{{\bf Q}}
\def\nv{\hat{\bf n}}
\def\plan{\hat{\bf \pi}}

\def\xu{\hat{\bf x}}
\def\yu{\hat{\bf y}}
\def\zu{\hat{\bf z}}

\def\tw{\tilde w}
\def\tv{\tilde v}
\def\lambdabar{\lambda\raise0.4ex\hbox{\kern-0.5em\hbox{--}}\ }
\def\lambdaC{\lambda\raise0.5ex\hbox{\kern-0.5em\hbox{--}}_{\rm C}}
\def\lesssim{{\lower0.5ex\hbox{$\stackrel{<}{\sim}$}}}
\def\gtrsim{{\lower0.5ex\hbox{$\stackrel{>}{\sim}$}}}
\begin{document}

%\title{The high-Z hydrogen-like atom : \\ a model for polarized structure functions}
%\maketitle

\centerline{THE HIGH-Z HYDROGEN-LIKE ATOM:} 

\medskip
\centerline{A MODEL FOR POLARIZED STRUCTURE FUNCTIONS\footnote{
Presented at $2^{nd}$ International Conference on Quantum Electrodyanmics and Statistical Physics (QEDSP2006), 19-23 September 2006, Karkhov, Ukraine
}}

\vskip1cm

\centerline{X. Artru\footnote{Universit\'e de Lyon;
Institut de Physique Nucl\'eaire de Lyon, CNRS and Universit\'e Lyon-I.
Domaine Scientifique de la Doua. 4, rue Enrico Fermi, F-69622 Villeurbanne, France. 
e-mail: x.artru@ipnl.in2p3.fr
}, 
K. Benhizia\footnote{
Laboratoire de Physique Math\'ematiques et Physique Subatomique, Universit\'e Mentouri, Constantine, Algeria
e-mail : Beni.Karima@laposte.net
}
}

\bigskip
 
\abstract
{
The Dirac equation offers a precise analytical description of relativistic two-particle bound states, when one of the constituent is very heavy and radiative corrections are neglected. Looking at the high-Z hydrogen-like atom in the infinite momentum frame and treating the electron as a "parton", various properties usually attributed to the quark distributions in the nucleon are tested, in particular: Bj\o rken scaling; charge, helicity, transversity and  momentum sum rules; existence of the parton sea; Soffer inequality; correlation between spin and transverse momentum (Sivers and Boer-Mulders effects); transverse displacement of the center-of-charge and its connection with the magnetic moment. Deep inelastic experiments with photon or positron beams at MeV energies, analogous to DIS or Drell-Yan reactions, are considered.
} 

\section{Theoretical frame}

The Dirac equation enables us to study the relativistic aspects of an hydrogen-like atom $A$ of large $Z$ ($Z \alpha \sim 1 $, where $\alpha=e^2/(4\pi)\simeq1/137$). It takes all orders in $Z \alpha$ into account but neglects (i) the nucleus recoil, (ii) the nuclear spin and (iii) radiatives corrections like the Lamb shift. So it is accurate at least to zero$^{th}$ order in $\alpha$ and $m_e/m_A$.
Applying a Lorentz boost, we have an explicit model of ``doubly relativistic'' two-body bound state (relativistic for the internal and external motions). In particular, boosting the atom to the ``infinite momentum frame'' (or looking it on the null-plane $t+z=0$), one has a model for the structure functions which appear in deep inelastic scattering on hadrons. In fact, since it neglects nucleus recoil, this model is best suited to mesons with one heavy quark. However many properties can be generalized to hadrons made of light quarks.

In analogy with the quark distributions, we introduce the unpolarized electron distributions $ q (k^+) $, $ q (\kv_T,k^+) $ and $ q (\bv, k^+ ) $ where $k^+$ takes the place of the Bjorken scaling variable, $\kv_T$ is the transverse momentum of the electron and the {\it impact parameter} $ \bv = (x,y)$ is the variable conjugate to $\kv_T$.
We will also define the corresponding polarized distributions like $q (\bv,k^+, \Sv^e  ; \Sv^A)$ where $\Sv^A$ and $\Sv^e$ are the polarization vectors of the atom and the electron. We will particularly study 
\begin{itemize}
\item
the differences between $q (k^+)$, the helicity distribution $\Delta q(k^+)$ and the transversity distribution $\delta q(k^+)$;
\item the sum rules for the vector, axial and tensor charges and for the longitudinal momentum;
\item the correlations between $\Sv^A$, $\Sv^e$ and $\bv$ or $\kv_T$, like the {\it Sivers effect};
\item the existence of a non-zero $\langle\bv\rangle$ for transverse $\Sv^A$ and its connection to the atom magnetic moment;
\item the positivity constraints;
\item the existence of an electron - positron sea and its role in the sum rules.
\end{itemize}
As scaling variable we take the null-plane momentum of the electron measured in the atom rest frame,
\begin{equation}
k^+ = \left(k_0 + k_z\right)_{rest\ frame} = M_A\ \left(k_z/P_{A,z} \right)_{inf.\ mom.\ frame} = M_A\ x_{Bj}
\label{def-k+}
\end{equation}
We prefer it to the Bj\o rken variable $x_{Bj}$ which is very small and depends on the nucleus mass. 
The kinematical limit for $| k^+ |$ is $ M_{atom} $ but typical values are $ | k^+ - m_e| \sim Z\alpha m_e $.  

We hope in this study to get a better insight of relativistic and spin effects in hadronic physics. The infinite momentum or null-plane descrition can also be interesting in atomic physics itself, since ``deep inelastic'' experiments can also be made with atoms, in particular:
\begin{itemize}
\item
Compton profile measurements: \par
$\gamma(K)+\ bound\ e^-\to \gamma(K+Q)+ \ free\ e^-(k')$,
\item
Moeller or Bhabha scattering: \par
$e^{\pm}(K)+\ bound\ e^-\to e^{\pm}(K+Q)+\ free\ e^-(k')$,
%(replacing the $\gamma$ by a $e^{\pm} $),  
\item
annihilation:
$e^+(K)+ \ bound\ e^-\to\gamma(K+Q)+\gamma(k)$. 
\end{itemize}
The Mandelstam variables $s=(K+Q+k')^2$, $t=Q^2$ and $u=(K-k')^2$ are supposed to be large compared to $m_e^2$. A ten MeV beam is sufficient for that. We take the $\zu$ axis opposite to the beam direction. In the laboratory frame the final particles are ultrarelativistic nearly in the $-\zu$ direction. The components $k^+$ and $\kv_T$ of the electron momentum just before the collision are given by  
\begin{equation}
k^+ \simeq Q^+
\,,
\label{k+simeq}
\end{equation}
\begin{equation}
\kv_T \simeq - \Pv'_T(nucleus) = \kv'_T + \Qv_T.
\label{kT} 
\end{equation}
$k^+$ can be measured with one detector, $k^+$ {\it and} $\kv_T$ need two detectors. The definition (\ref{kT}) of $\kv_T$ is ambiguous due to the final state Coulomb interaction.

\section{Joint $(\bv,k^+)$ distribution}
Being observable quantities, the operators $k^+$ and $\kv_T$ should be defined in the gauge independent way
\begin{equation}
k^+ = i\partial_0-V(\rv) -i\partial_z-A_z(\rv),
\qquad
\kv_T = -i\nabla_T(\rv)-\Av_T.
\label{mecanique}
\end{equation}
They do not commute: $[k^+,\kv_T]= -i \nabla_T V(\rv)$, where $-\nabla_T V(\rv)$ is the transverse part of the Coulomb force. Therefore one cannot define a joint distribution $q(k^+,\kv_T)$ in an unambiguous way. Leaving this problem for the next section, we can at least define the joint distribution $q(k^+,\bv)$ in the impact parameter representation. This quantity plays a role in double H + H collisions in which both nuclei and both electrons collide. 

From the known the Dirac wave function of the hydrogen atom \cite{Bjorken}
\begin{equation}
\Psi(t,\rv) = \Psi(\rv) \ e^{-iEt},
\label{Psi}
\end{equation}
we can define the two-component null-plane wave function in $\bv$ and $\kv_T$ \cite{Como},
\begin{equation}
\Phi (\bv,k^+) =  \int_{-\infty}^{+\infty} dz \ \exp\left\{-i k^+ z +i E z -i \chi(\bv,z) \right\} \ \Phi (\rv ), 
\label{Phibk+}
\end{equation}
\begin{equation}
\Phi=(\rv) =\pmatrix{
\Psi_1(\rv)+\Psi_3(\rv) \cr 
\Psi_2(\rv)-\Psi_4(\rv)}\,,
\label{Phi}
\end{equation}
\begin{equation}
\chi(\bv,z) = \int_{z_0}^z dz' \ V(x,y,z') 
= -Z\alpha \ \left[ \sinh^{-1} \left( {z \over b} \right) - \sinh^{-1} \left( {z_0 \over b} \right) \right].
\label{chi}
\end{equation}
The "gauge link" $\exp\{-i\chi(\bv,z)\}$ transforms $\Psi$ in the Coulomb gauge to $\Psi$ in the null plane gauge $A^+=0$ ($A_z=0$ in the infinite momentum frame). The choice of $z_0$ corresponds to a residual gauge freedom. The quantity
\begin{equation}
q(\bv,k^+)\equiv {dN_{e^-}\over d^2\bv \ dk^+ / (2\pi)} = \ \Phi^\dagger(\bv,k^+) \ \Phi(\bv,k^+),
\label{dens(b)} 
\end{equation}
will be temporarily interpreted as the electron distribution in the atom. One has indeed
\begin{equation}
\int_{-\infty}^{+\infty} {dk^+\over2\pi} \ q(k^+) = \int_{-\infty}^{+\infty} {dk^+\over2\pi} \int d^2\bv \ q(\bv,k^+) = 1.
\label{norma}
\end{equation}
However a significant re-interpretation will be given in Section 5.

The gauge link makes $q(k^+,\bv)$ invariant under a gauge transformation, for instance $V(\rv) \to V(\rv) +$ Constant, $E \to E +$ Constant. Such a shift of the potential is practically realized when electrons are added in far outer shells. Intuitively, this addition does not change the momentum distribution of the deeply bound electrons. 

\section{Joint $(\kv_T,k^+)$ distribution}

Notwithstanding the non-commutativity mentioned earlier, one can make a transversal Fourier transform of (\ref{Phibk+}),  
\begin{equation}
\Phi(\kv_T,k^+) %= \int d^3 \rv \ e^{-i \kv_T\cdot\bv - ik^+ z -i \chi(\rv)} \ \Phi(\rv) 
= \int d^2 \bv \ e^{-i\kv_T\cdot\bv}\ \Phi(\bv,k^+)
\label{b->kT}
\end{equation}
and define a longitudinal-transverse momentum distribution
\begin{equation}
q ( \kv_T , k^+ ) = \Phi^\dagger ( \kv_T , k^+ )  \ \Phi ( \kv_T , k^+ ), 
\label{long-transv}
\end{equation}
normalized to
\begin{equation}
q ( k^+ ) = \int q ( \kv_T , k^+ ) \ {d^2\kv_T / (2 \pi)^2}.
\end{equation}
$q ( \kv_T , k^+ )$ depends on $z_0$, which is a remnent of the ambiguity. However, an appropriate choice of $z_0$ turns this apparent disease into an advantage \cite{Brodsky,Yuan,Collins}. Taking $z_0=-\infty$ for the Compton reaction, the factor $e^{-i\chi(\bv,z)}$ in (\ref{Phibk+}) just takes care of the final state interaction: it describes the distortion of the scattered electron wave function by the Coulomb potential, in the eikonal approximation. Similarly, taking $z_0=+\infty$ for the annihilation reaction, it describes the distortion of the initial positron wave function.
Thus $q(\kv_T,k^+)$, which depends on $z_0$, has no precise intrinsic character. One can just consider a ``most intrisic'' definition with $z_0=0$.

\section{Spin dependence of the electron density}

In formulas (\ref{Psi}-\ref{long-transv}) the angular momentum state of the atom was not specified.
We assume that the electron is in the fundamental $n=1$, $j=1/2$ state and the nucleus is spinless. 
Let $\Sv^A = 2 \langle \jv \rangle $ and $\Sv^e = 2 \langle \sv \rangle $ denote the atom and electron polarization vectors. The {\it unpolarized} electron density in $(\bv,k^+ )$ space in a fully polarized ($|\Sv^A|=1$) atom is 
\begin{equation}
q (\bv,k^+ ; \Sv^A ) \ = \ 
\Phi^\dagger(\bv,k^+ ; \Sv^A ) \ \ \Phi(\bv,k^+ ; \Sv^A )
\label{dens}
\end{equation}
and the electron polarisation is given by
\begin{equation}
\Sv^e (\bv,k^+ ; \Sv^A ) \ \ q (\bv,k^+ ; \Sv^A ) \ = \  
\Phi^\dagger(\bv,k^+ ; \Sv^A ) \ \ \vec\sigma \ \ \Phi(\bv,k^+ ; \Sv^A )
\,.
\label{spin-dens}
\end{equation}
Taking into account parity and angular momentum conservations, the density of electrons with polarization $\Sv^e$ in a polarized atom can be written as
$$%\begin{equation}
q (\bv,k^+,\Sv^e;\Sv^A) = {q(b,k^+)\over2} [1+ 
C_{0n} \ \Sv^A  \cdot \nv +
C_{n0} \ \Sv^e \cdot \nv +
C_{nn} \ (\Sv^e \cdot \nv)  ( \Sv^A  \cdot \nv) 
$$%\nonumber\end{equation}
\begin{equation}
+ \ C_{ll} \  S^e_z  S^A_z  +
C_{l \pi } \ S^e_z  (\Sv^A  \cdot \plan) \ +
C_{ \pi l} \ (\Sv^e \cdot \plan) S^A_z  +
C_{\pi\pi} \ (\Sv^e \cdot \plan) (\Sv^A  \cdot \plan) ],  
\label{densPol}
\end{equation}%
where $\plan = \bv / b $ and $\nv = \zu \times \plan $. 
%form a basis of transverse vectors.
The $ C_{i,j} $'s also are functions of $ b $ and $k^+$. A similar equation can be written in the $\kv_T$ representation.
%~\cite{Bacchetta-BHM}
Integrating (\ref{densPol}) over $\bv$ leaves the following spin correlations:
\begin{equation}
q (k^+, \Sv^e  ; \Sv^A)  
= {1\over2} \ \left[ \ q (k^+) \ + \ \Delta q (k^+) \  S^e_z S^A_z  
\ + \ \delta q (k^+) \ \ \Sv^e_T \cdot \Sv^A_T \ \right]. 
\label{k+distrib}
\end{equation}
where $\Delta q(k^+)$ and $\delta q(k^+)$ are the helicity and transversity distributions.

\subsection{ Formulas for the polarized densities in $(\bv,k^+)$ and $(\kv_T,k^+)$}

For the $ j_z = + 1/2 $ state, $\Phi$ of Eqs.(\ref{Phibk+}-\ref{b->kT}) can be written as
\begin{equation} 
\Phi(\bv, k^+ ; \Sv^A=+\zu) = 
\left( 
\begin{array}{c}
w \\ 
-iv e^{i\phi} \\ 
\end{array}
\right),\qquad
\Phi(\kv_T, k^+; \Sv^A=+\zu) = 
\left( 
\begin{array}{c}
\tw \\ 
- \tv e^{i\phi} \\ 
\end{array}%
\right).
\label{Phi-up} 
\end{equation}
For the $ j_z = - 1/2 $ state,
\begin{equation} 
\Phi(\bv, k^+ ; \Sv^A=-\zu) = 
\left( 
\begin{array}{c}
iv e^{-i\phi}  \\ 
w \\ 
\end{array}
\right),\qquad
\Phi(\kv_T, k^+; \Sv^A=-\zu) = 
\left( 
\begin{array}{c}
\tv e^{i\phi} \\ 
 \tw \\ 
\end{array}%
\right).
\label{Phi-down} 
\end{equation}
For other orientations of $\Sv^A$, one takes linear combinations of (\ref{Phi-up}) and (\ref{Phi-down}). 
The $(\bv, k^+)$ distribution depends of $z_0$ only in an over-all phase. Chosing $z_0=0$, $v(\bv,k^+)$ and $w(\bv,k^+)$ are real and given by
\begin{equation}
\pmatrix{ v \cr w} = \int_{-\infty}^{\infty } dz \ 
\pmatrix{ \xi b  \cr r + i \xi z }
\ e^{-i k^+ z +i E z -i \chi(\bv,z)} \ f(r)/r, 
\label{v-w}
\end{equation}
where $\xi = Z\alpha /(1+\gamma)$, $\ \gamma = E/m_e = \sqrt{1-(Z\alpha)^2}$ and
\begin{equation}
f(r) = \left( {1+\gamma\over8\pi\Gamma(1+2\gamma)}\right)^{1/2} \ (2m_eZ\alpha)^{\gamma+1/2} 
\quad r^{\gamma - 1} \ \exp(-m_e Z\alpha r)
\end{equation}
is the 1S radial wave function. Then,
\begin{eqnarray}
q(b, k^+) &=& w^2 + v^2  \nonumber \\
C_{nn} (b, k^+) &=& 1 \nonumber  \\
C_{0n} (b, k^+)  = C_{n0} (b, k^+)  
&=& - 2 \ wu / (w^2 + v^2 ) \nonumber  \\
C_{ll} (b, k^+)  = C_{\pi \pi } (b, k^+) 
&=& ( w^2 - v^2 ) / (w^2 + v^2 ) \nonumber \\
C_{l \pi} (b, k^+)  = C_{ \pi l} (b, k^+) &=& 0.
\label{Cij(b)}
\end{eqnarray}
Note that $C_{0n}(b,k^+)\ne0$, which gives an asymmetrical impact parameter profile for a transversely polarized atom.

The $(\kv_T, k^+)$ distribution depends on $z_0$. Taking $z_0 = \mp \infty $ makes (\ref{chi}) diverges. 
In practice we will assume that $|z_0|$ is large but finite, accounting for a screening of the Coulomb potential. It gives 
\begin{equation}
\chi(\bv,z) = 
-Z\alpha \ \left[\ \sinh^{-1} (z/b) - \epsilon(z_0) \ \ln(2 |z_0|/b)\ \right],
\end{equation}
with $\epsilon(0)=0$ and $\epsilon(\mp\infty)=\mp1$, the upper sign corresponding to Compton scattering and the lower sign to annihilation. Modulo an overall phase, 
\begin{equation}
\pmatrix{\tw \cr \tv}
= 2\pi \int_0^\infty b \ db \ b^{iZ\alpha \epsilon(z_0)} \  
\pmatrix{ 
J_0 (k_T b ) \ w (b,k^+) \cr
J_1 (k_T b ) \ v (b,k^+)}.
\end{equation}
The analogue of (\ref{Cij(b)}) is
\begin{eqnarray}
q(k_T,k^+) &=& |\tw|^2 + |\tv|^2  
\nonumber \\
C_{nn} (k_T , k^+) &=& 1 \nonumber  \\
C_{0n} (k_T , k^+)  = C_{n0} (k_T , k^+)  
&=& 2 \Im (\tv^*\tw) / (|\tw|^2 + |\tv|^2 )  
\nonumber \\
C_{ll} (k_T , k^+)  = C_{\pi \pi } (k_T , k^+) 
&=& ( |\tw|^2 - |\tv|^2 ) / (|\tw|^2 + |\tv|^2 ) \nonumber  \\
C_{l \pi} (k_T , k^+)  = - C_{ \pi l} (k_T , k^+) 
&=& 2 \Re (\tv^*\tw) / (|\tw|^2 + |\tv|^2 ) 
\,. 
\label{Cij(k)}
\end{eqnarray}
These coefficients are related to the structure functions listed in Ref.\cite{Boer-Mulders}.
For the ``most intrisic'' gauge $z_0=0$, $\tw$ and $\tv$ are real so that $C_{0n}(k_T,k^+)=0$ (no Sivers effects). This is in accordance with time reversal invariance \cite{Collins}. For the ``Compton'' and ``annihilation'' gauges ($z_0=\mp1$), $\tw$ and $\tv$ are complex numbers, so that Sivers \cite{Sivers} effect ($C_{0n}(k_T,k^+)\ne0$ and the Boer-Mulders \cite{Boer-Mulders} effect ($C_{n0}(k_T,k^+)\ne0$) take place.

In the Compton case the factor $ b^{- iZ\alpha} $ behaves like a converging cylindrical wave. Multiplying $\Phi(\rv)$, it operates as a boost toward the $z$ axis, interpreted as the "focusing" of the final particle by the Coulomb field~\cite{Burkardt}. This focusing converts the asymmetry in $\bv$ for a transversely polarized atom into the Sivers asymmetry in $\kv_T$. 
The opposite effect (defocusing of the positron) takes place in the annihilation case. 

\subsection{Sum rules }
Integrating (\ref{k+distrib}) over $k^+$, one obtains the vector, axial and tensor charges
\begin{equation}
q = \int d^3 \rv \ \Psi^\dagger(\rv;\Sv^A) \ \Psi(\rv;\Sv^A) = 1,
\label{V-charge}
\end{equation}
\begin{equation}
\Delta q = \Sv^A\cdot \int d^3 \rv \ \Psi^\dagger(\rv ; \Sv^A ) \ \vec\Sigma \ \Psi(\rv ; \Sv^A)
= {1- \xi^2 /3 \over 1 + \xi^2},
\label{A-charge}
\end{equation}
\begin{equation}
\delta q = \Sv^A\cdot \int d^3 \rv \ \Psi^\dagger(\rv ; \Sv^A ) \ \beta \ \vec\Sigma \ \Psi(\rv ; \Sv^A)
= {1+ \xi^2 /3 \over 1 + \xi^2}.
\label{T-charge}
\end{equation}
Note the big "helicity crisis", $\Delta q=1/3$ instead of 1 as naively expected, for $Z\alpha = 1$.

\subsubsection{Sum rule for the atom magnetic moment}
Consider a classical object of mass $M$, charge $Q$, spin $\Jv$ and time-averaged magnetic moment $\vec\mu$ in its rest frame. In this frame, the centre of energy $\rv_G$ and the average center of charge $\langle \rv_C \rangle$ coincide, say at $\rv = 0$. Upon a boost of velocity $\vv$, $\ \rv_G$ and $\langle \rv_C \rangle$ undergo the lateral displacements 
\begin{equation}
\bv_G = \vv \times \Jv /M
\,,\quad 
\langle \bv_C \rangle = \vv \times \vec \mu /Q
\,.
\end{equation}
$\bv_G $ and $ \langle \bv_C \rangle $ coincide if the gyromagnetic ratio 
has the Dirac value $Q/M$. In our case, $ \bv_G $ is negligible due to the large nucleus mass, therefore the magnetic moment is almost totally anomalous. In the infinite momentum or null-plane frame ($\vv \simeq \zu$) one observes an electric dipole moment~\cite{Burkardt} 
\begin{equation}
-e \ \langle \bv \rangle  
= \mu_{atom} \ \zu \times \Sv^A, 
\end{equation}
which we can calculate from $C_{0n}(b,k^+)$. Weighting (\ref{dens}) with $b_x\equiv x$ for $\Sv^A=\yu$ one obtains 
\begin{equation}
\langle x \rangle = - (1+2\gamma)/(6m_e)
\end{equation}
which is accordance with the relativistic result for the atomic magnetic moment $\mu_A=-e\ (1+2\gamma)/(6m_e)$ (ignoring the anomalous magnetic moment of the electron itself).

\subsection{\protect\bigskip Positivity constraints}

The spin correlations between the electron and the atom can be encoded in a positive-definite "grand density matrix"  $R$~\cite{Artru-Richard}, 
\begin{equation}
R = C_{\mu\nu} \ \sigma^\mu_A \otimes \left[\sigma^\nu_e\right]^t.
\label{decompose-R}
\end{equation}
Here $\mu$ and $\nu$ run from 0 to 3, summation is understood over repeated indices, $\sigma^0 = I$ and $C_{00}=1$. 
$R$ can be seen as the density matrix of the final state in the crossed reaction 
$nucleus \to atom(\Sv^A) + e^+(-\Sv^e) $. Besides the trivial conditions $| C_{ij} | \le 1$ the positivity of $R$ gives
\begin{equation}
( 1 \pm C_{nn} )^2  \ge 
(  C_{n0}  \pm C_{0n}  )^2 +
(  C_{ll} \pm  C_{\pi\pi} )^2 +
(  C_{\pi l}  \mp  C_{l\pi} )^2.
\label{cond}
\end{equation}
These two inequalities agree with those of Ref.\cite{Bacchetta}. Together with $| C_{ll} | \le 1$ they are saturated by (\ref{Cij(b)}) or (\ref{Cij(k)}). This maximal strength of the spin correlation means that the information contained in the atom polarization is fully transferred to the electron, once the other degrees of freedom ($k^+$ and $\bv$ or $\kv_T$) have been fixed. If we integrate over $k^+$, for instance, some information is lost and some positivity conditions get non-saturated. The same happens if there are ``spectators'' electrons which keep part of the information for themselves.

After integration over $\bv$ or $\kv_T$, we are left with the 
{\it Soffer inequality} \cite{Soffer}, 
\begin{equation}
2 \ | \delta q (k^+) | \le q (k^+) + \Delta q (k^+), 
\label{Soffer}
\end{equation}
which are saturated by (\ref{V-charge}-\ref{T-charge}).
 
Note that a complete anti-correlation between the atom and the electron spins, $C_{ll}=C_{nn}=C_{\pi\pi}=-1$ and $C_{i\ne j}=0$, leading to
$$
q(\bv,k^+,\Sv^e;\Sv^A)= q(\bv,k^+)\ (1-\Sv^e\cdot\Sv^A)/2,
$$ 
violates the positivity conditions, although the last expression is positive for any $\Sv^e$ and $\Sv^A$. In fact such a correlation would make $\langle A,e^+|R|A,e^+\rangle$ negative for some {\it entangled} states $|A,e^+\rangle$ in the crossed channel, in particular the spin-singlet state \cite{Artru-Richard}. 

\section{\protect\bigskip The electron-positron sea }
The charge rule (\ref{V-charge}) receives positive contributions from both positive and negative values of $k^+$.   
So the contribution of the positive $k^+$ domain is less than unity. On the other hand, physical electrons have positive $k^+$. It seems therefore that there is less than one physical electron in the atom. This paradox is solved by the second quantification and the introduction of the electron-positron sea. 

Let us denote by $|n\rangle$ an electron state in the Coulomb field. Quantizing the states in a box, we take $n$ to be integer. Negative $n$'s are assigned to negative energy states. Positive $n$'s up to $ n_B$ label the bound states ($-m_e<E_n<+ m_e$) and the remaining ones from $ n_B+1$ to $ + \infty $ label the positive energy scattering states, $E_n\ge+m_e$.
Let $ | k,s \rangle $ be the plane wave with four-momentum $k$ and spin $s$, solution of the {\it free} Dirac equation. 
The destruction and creation operators in the interacting and free bases are related by
\begin{equation}
\alpha_{k,s}  = \sum_n \ \langle k,s | n \rangle \ a_n 
\,,\qquad
\alpha^\dagger_{k,s}  = \sum_n \ a^\dagger_n \ \langle n | k,s \rangle. 
\label{passage}
\end{equation}
In the Dirac hole theory, the hydrogen-like atom is in the Fock state
\begin{equation}
| H_n \rangle = a^\dagger_n \ a^\dagger_{-1} \ a^\dagger_{-2} 
\cdots a^\dagger_{-\infty} 
\ | \hbox{Dirac-bare nucleus} \rangle. 
\label{Fock}
\end{equation}
``Dirac-bare'' means that all Dirac states, including the negative energy ones, are empty. The number of electrons of momentum $k$ and spin $s$ (with positive $k^0$ and $k^+$) in the atom is 
\begin{equation}  
N^{e^-}_{atom} (\kv,s) = \langle \alpha^\dagger_{k,s} \ \alpha_{k,s} \rangle_{atom}  
= | \langle k,s | n \rangle  |^2 
+ \sum_{n'<0}| \langle k,s | n' \rangle  |^2.
\label{elec/atom}
\end{equation}
A stripped ion is a "Dirac-dressed" nucleus, all negative energy states being occupied. For the ion the factor $a^\dagger_n$ of (\ref{Fock}) is missing and the first term of (\ref{elec/atom}) is absent. By difference,
\begin{equation}  
N^{e^-}_{atom} - N^{e^-}_{ion} = \sum_{k, k^0>0} \ \sum_s 
\ | \langle k,s | n \rangle  |^2 
= \int_{k^+ >0} {dk^+ \over 2 \pi} \ q ( k^+ ),
\label{k+pos}
\end{equation}
the last expression being for the continuum limit $\langle k,s|n\rangle\ \longrightarrow\ \Phi(\kv_T,k^+ )$.

Positrons are holes in the Dirac sea. The number of positrons of momentum $\bar k$ (with positive $\bar k^0$ and $\bar k^+$) and spin $\bar s$ in the atom is 
\begin{equation}  
N^{e^+}_{atom} (\bar\kv,\bar s) = 
\langle \alpha_{-\bar k,-\bar s} \ \alpha^\dagger_{-\bar k,-\bar s}  \rangle_{atom}
= \sum_{0 < n' \ne n}| \langle -\bar k,-\bar s | n' \rangle  |^2
\,.
\label{posit/atom}
\end{equation} 
For the ion, the condition $n' \ne n$ is relaxed. By difference, 
\begin{equation}
N^{e^+} _{ion} - N^{e^+} _{atom} 
= \int_{k^+ < 0} {dk^+ \over 2 \pi} \ q(k^+), 
\label{posit/at-nuc}
\end{equation}
where we have made the change of variable $k=-\bar k$. The sum rule (\ref{V-charge}) can therefore be interpreted in the following way:
\begin{itemize}
	\item for $k^+>0$, $\ q(k^+) =$ %dN^{e^-}_{atom}/dk^+ - dN^{e^-}_{ion}/dk^+ 
	($e^-$ distrib. in atom) - ($e^-$ distrib. in ion)
	\item for $k^+<0$, $\ q(k^+) =$ % dN^{e^+}_{ions}/dk^+ - dN^{e^+}_{atom}/dk^+$
	($e^+$ distrib. in ion) - ($e^+$ distrib. in atom)
\end{itemize}
Thus
\begin{equation}
\left(N^{e^-}_{atom} - N^{e^-}_{ion} \right) + \left(N^{e^+} _{ion} - N^{e^+} _{atom}\right) = 1,
\end{equation}
where each braket $\in[0,1]$. Introducing $Q^e = N^{e^-}\!-\! N^{e^+}$, this can be rewritten as
\begin{equation}
Q^e_{atom} - Q^e_{ion}= 1
\end{equation}
$Q^e_{ion}$ is the electronic charge renormalisation of the ion on the null plane. It is more likely positive, maybe infinite for a pointlike nucleus. The renormalisation $Q^e_{atom}-1$ of the atom is equal to it. It may be interesting to relate $Q^e_{ion}$ with the result of covariant QED. 
%Eqs.(\ref{k+pos}) + (\ref{posit/at-nuc}) and (\ref{V-charge})

$dN^{e^-}_{atom}/dk^+$ , $dN^{e^+}_{atom}/dk^+$ , $dN^{e^-}_{ion}/dk^+$ and $dN^{e^+}_{ions}/dk^+$ are separately measured in the deep inelastic reactions listed in Section 1 and their generalizations to the electron - positron sea, for instance 
\begin{equation}
\gamma(K)+\ sea\ e^\pm\to \gamma(K+Q)+ \ free\ e^\pm(k').
\label{sea}
\end{equation} 
A ``sea'' electron can be equally understood in the sense given by Feyman in the parton model or by Dirac in the hole theory. It gives the second term of (\ref{elec/atom}) and the whole process is
\begin{equation}
\gamma+A\to\gamma'+A +slow\ e^+ +fast\ e^-. 
\label{sea-}
\end{equation} 
A ``sea'' positron is understood in the Feynman sense only. In the hole theory, an {\it electron of large negative energy} is lifted to a bound or slow free state $|n'\rangle$. It gives the right-hand side of (\ref{posit/atom}). The whole process is
\begin{equation}
\gamma + H_n\to H^-_{n,n'}+fast\ e^+ +\gamma'.
\label{sea+}
\end{equation}
Of course, one can permute the roles of the electron and the positron in the Dirac theory; then Feynman and Dirac sea positrons become equivalent. 

\section{Momentum sum rule}

$q(k+)$ obeys a momentum sum rule which, like the charge sum rule, applies to the difference between the atom and the ion. 

The null-plane momentum of the ion (=nucleus) can be decomposed into a matter part and a Coulomb field part:  
\begin{equation}
P^+_{ion} =M_N = P^+_{matter\,N} + P^+_{field}\left\{\Ev_N\right\}.
\label{M_ion}
\end{equation}
$P^+_{matter\,N}$ includes the momentum of the electron cloud which renormalize the ion charge.
$P^+_{field}\left\{\Ev_N\right\}$ is the flux of the $T^{+,\nu}=T^{0\nu}+T^{z\nu}$ component of the energy-momentum tensor $T^{\mu\nu}\left\{\Ev_N\right\}$ of the nucleus Coulomb field $\Ev_N(\rv)$ across the null plane:
\begin{eqnarray}
P^+_{field}\left\{\Ev_N\right\}
&=& \int \left(T^{0\nu}\left\{\Ev_N\right\}+T^{z\nu}\left\{\Ev_N\right\}\right) \ d\sigma_\nu\\
&=& \int dx\,dy\,dz\ \left(T^{00}+T^{0z}+T^{z0}+T^{zz}\right)\\
&=& \int dx\,dy\,dz\ \left(E_x^2 + E_y^2\right).
\label{k+coul}
\end{eqnarray}
We have used $d\sigma_\nu=(1,0,0,1)\,dx\,dy\,dz$, $T^{00}={1\over2}\left(\Ev^2+\Bv^2\right)$, $T^{zz}=T^{00}-E_z^2-B_z^2$, $T^{0z}=T^{z0}=E_xB_y-E_yB_x$. Similarly, for the atom, we have 
\begin{equation}
P^+_{atom}=M_A = P^+_{matter\,N} + P^+_{bare\ e} + P^+_{field}\left\{\Ev_N+\Ev_e\,;\,\Bv_e\right\}.
\label{M_atom}
\end{equation}
the electron magnetic field being included. Here $P^+_{bare\ e}$, $\Ev_e$ and $\Bv_e$ take only into account the difference between the atomic and ionic electron clouds. Substracting (\ref{M_ion}) from (\ref{M_atom}),
\begin{equation}
E\equiv M_A-M_N = P^+_e + P^+_{int}.
\label{interfer}
\end{equation}
$P^+_{int}$ results from the crossed terms in $\Ev_N$ and $\Ev_e$ or $\Bv_e$ of $T^{\mu\nu}$. Its value is
\begin{eqnarray}
P^+_{int}&=&2\int d^3\rv\ \left[\ E_{Nx}\left(E_{ex}+B_{ey}\right) + E_{Ny}\left(E_{ey}-B_{ex}\right)\ \right] 
\nonumber\\
&=& {4\over3}\int d^3\rv\ \Ev_N\cdot \Ev_e 
= {4\over3}\langle V\rangle.
\label{P+int}
\end{eqnarray}
$\langle V\rangle$ is the average potential energy. The terms in $\Ev_N \Bv_e$ have disappeared upon angular integration.

$P^+_e= P^+_{bare\ e} + P^+_{field}\left\{\Ev_e\right\}$ is the mean value of the null-plane mechanical momentum $k^+$ of the physical electron (more precisely the ``atom - minus - ion'' part of it). Inserting $k^+=E-i\partial_z$ in Eqs.(\ref{dens(b)}), one obtains
\begin{equation}
P^+_e=\int_{-\infty}^{+\infty} k^+ \ {dk^+\over2\pi} \ q(k^+) = E-{4\over3}\langle V\rangle
\label{k+somme}
\end{equation}
with 
\begin{equation}
\langle V\rangle=\int d^3\rv \ \Psi^\dagger(\rv) \ V(r)\ \Psi(\rv) = -m_e(Z\alpha)^2/\gamma. 
\label{<V>}
\end{equation}
Eqs.(\ref{interfer}-\ref{k+somme}) constitute the momentum sum rule.

\section{CONCLUSION}

This study has shown the rich spin and $\kv_T$ structure of the hydrogen-like atom at large $Z$ when it is observed in the infinite momentum (or null-plane) frame. Without the complications of QCD, like gluon self-interaction and confinement, many properties attributed to the leading twist hadronic structure functions have been found and clearly interpreted here, in particular: the sum rules, the spin crisis, the connection between $\langle \bv \rangle\ne0$ and the Sivers effect, the relation between $\langle \bv \rangle$ and the magnetic moment, the role of spectators in the positivity constraints, the existence of a Feynman sea.
 With this "theoretical laboratory" one may also investigate non-leading twist structure functions, elastic form factor {\it a la} Isgur-Wise, etc. Our results are interesting also in pure QED. We have seen a connection between the nucleus charge renormalization and the unpolarized deep inelastic structure function of the $e^\pm$ cloud of a stripped ion target. Thus the charge renormalization can be analyzed experimentally in deep inelastic Compton, M\o ller or annihilation processes. The same reactions at the 10 MeV energy scale can test the relativistic corrections to the electronic wave functions of large $Z$ atoms. 

The powerpoint document presented at QEDSP-2006, which contains figures not presented here, can be obtained upon request to one author (X. A.).

 %\begin{center}
%\setcounter{page}{12}{\Large \addcontentsline{toc}{chapter}{R��ences}}
%\end{center}

\end{document}